\documentclass[11pt]{article}

\makeatletter\renewcommand{\section}{\@startsection
{section}{1}{\z@}{-3.5ex plus -1ex minus
    -.2ex}{2.3ex plus .2ex}{\bf }}

\makeatletter\renewcommand{\subsection}{\@startsection{subsection}{2}{\z@}{-3.25ex
plus -1ex minus
   -.2ex}{1.5ex plus .2ex}{\it }}
\makeatletter\renewcommand{\subsubsection}{\@startsection{subsubsection}{3}{-2.45ex}{-3.25ex
plus -1ex minus -.2ex}{1.5ex plus .2ex}{\it }}
\renewcommand{\thesection}{\arabic{section}}
\renewcommand{\thesubsection}{\arabic{section}.\arabic{subsection}.}

\renewcommand{\theequation}{\thesection.\arabic{equation}}
\makeatletter \@addtoreset{equation}{section}

\usepackage{graphicx}
\usepackage{epsfig} 

\newcommand{\be}{\begin{equation}}
\newcommand{\ee}{\end{equation}}
\newcommand{\bea}{\begin{array}}
\newcommand{\ea}{\end{array}}
\newcommand{\beqa}{\begin{eqnarray}}
\newcommand{\eeqa}{\end{eqnarray}}
\newcommand{\nn}{\nonumber}

\renewenvironment{thebibliography}[1]
     {\baselineskip=16pt plus 2pt minus 1pt
      \section*{\large\refname
        \@mkboth{\MakeUppercase\refname}{\MakeUppercase\refname}}%
     \list{\@biblabel{\@arabic\c@enumiv}}%
           {\settowidth\labelwidth{\@biblabel{#1}}%
            \leftmargin\labelwidth
            \advance\leftmargin\labelsep
            \@openbib@code
            \usecounter{enumiv}%
            \let\p@enumiv\@empty
            \renewcommand\theenumiv{\@arabic\c@enumiv}}%
      \sloppy
      \clubpenalty4000
      \@clubpenalty \clubpenalty
      \widowpenalty4000%
      \sfcode`\.\@m}

\let\fn\footnote
\renewcommand{\footnote}[1]{\linespread{1.1}\fn{#1}\linespread{1.29}}

\usepackage{amsfonts}
\usepackage{amssymb}
\usepackage{mathrsfs}
\usepackage{amsmath,amssymb}
\usepackage{bbm}
\usepackage{bm}

\newcommand{\appendices}{\section*{Appendix}\setcounter{subsection}{0} \setcounter{equation}{0}
\renewcommand{\thesubsection}{\Alph{subsection}.}
\renewcommand{\theequation}{\thesubsection\arabic{equation}}
}

\def\tyng(#1){\hbox{\tiny$\yng(#1)$}}

\begin{document}

\begin{titlepage}
\begin{flushright}
ITP-UH-06/09\\
\end{flushright}
\vskip 2.0cm

\begin{center}

\centerline{{\Large \bf Equivariant Reduction of
    Yang-Mills Theory }} 
\vskip 1em
\centerline{{\Large \bf over the Fuzzy Sphere and the Emergent Vortices}}

\vskip 2em

\centerline{\large \bf Derek Harland and Se\c{c}kin~K\"{u}rk\c{c}\"{u}o\v{g}lu}

\vskip 2em

{\small \centerline{\sl Institut f\"ur Theoretische Physik,
Leibniz Universit\"at Hannover} \centerline{\sl Appelstra\ss{}e 2,
D-30167 Hannover, Germany}

\vskip 1em

{\sl  e-mails:}  \hskip 2mm {\sl harland@math.uni-hannover.de \,, seckin.kurkcuoglu@itp.uni-hannover.de} }
\end{center}

\vskip 2cm

\begin{quote}
\begin{center}
{\bf Abstract}
\end{center}
\vskip 2em

We consider a ${\rm U(2)}$ Yang-Mills theory on ${\cal M} \times S_F^2$
where ${\cal M}$ is a Riemannian manifold and $S_F^2$ is the fuzzy
sphere. Using essentially the representation theory of ${\rm SU(2)}$ we 
determine the most general ${\rm SU(2)}$-equivariant gauge field on
${\cal M} \times S_F^2$. This allows us to reduce the Yang-Mills theory on
${\cal M} \times S_F^2$ down to an abelian Higgs-type
model over ${\cal M}$. Depending on the enforcement (or non-enforcement) 
of a ``constraint'' term, the latter may (or may not) lead
to the standard critically-coupled abelian Higgs model in the commutative limit,
 $S_F^2 \rightarrow S^2$.  For ${\cal M} = {\mathbb R}^2$, we find that  
the abelian Higgs-type model admits vortex solutions corresponding to 
instantons in the original Yang-Mills theory.  Vortices are in 
general no longer BPS, but may attract or repel according to the 
values of parameters.

\vskip 5pt

Pacs: 11.19.Nx, 11.10.Kk, 11.27.+d, 11.30.Ly
\end{quote}

\vskip 1cm
\begin{quote}
\end{quote}

\end{titlepage}

\setcounter{footnote}{0}

\newpage

\section{Introduction}

It is commonplace in modern physics to consider field theories defined on manifolds of the form ${\cal M}\times X$, where ${\cal M}$ represents physical space and $X$ is some compact manifold.  One popular example is to consider pure Yang-Mills theory, with $X$ a coset space $G/H$.  In this case the group $G$ acts naturally on its coset; by requiring the gauge fields to be invariant under the action of $G$ up to a gauge transformation, one obtains a new gauge theory on ${\cal M}$.  In this way a relatively complicated theory on ${\cal M}$ is obtained from a relatively simple theory on ${\cal M}\times X$.  We shall call such a process ``equivariant reduction''.

The first example of equivariant reduction was due to Witten \cite{Witten}.  He showed that Yang-Mills theory on $\mathbb{R}^4$ reduces under ${\rm SU(2)}$-equivariance to an abelian Higgs model on a 2-dimensional hyperbolic space $\mathbb{H}^2$, and thereby constructed the first instantons with charge greater than 1.  The space $\mathbb{H}^2$ emerges naturally in this example, because $\mathbb{R}^4\backslash \mathbb{R}^2$ is conformal to $\mathbb{H}^2\times S^2$, and Yang-Mills theory is conformally invariant in four dimensions.

In subsequent years two major formalisms have been developed to perform more exotic equivariant reductions.  Historically, the first was ``coset space dimensional reduction'' (CSDR) \cite{Forgacs,Zoupanos}, which uses intrinsic coordinates on the coset space, and is generally used as a method to try to obtain the standard model on the Minkowski space ${\cal M} = M^4$ starting from a Yang-Mills-Dirac theory on the higher dimensional space $M^4 \times G/H$. The second is the ``quiver'' approach \cite{Lechtenfeld:2003cq, Popov-Szabo, Popov, Dolan-Szabo}, which uses a more sophisticated language of equivariant vector bundles, and has the interesting feature of reducing self-dual instantons on ${\cal M}\times X$ to BPS vortices on ${\cal M}$.  The two approaches seem on the whole to be equivalent, but tend to emphasise different features of equivariant reduction.  In particular, Witten's example is the basic one in both approaches.

The quiver approach has also been applied to the case where ${\cal M}$ is a non-commutative manifold (the $2d$-dimensional  Moyal space ${\mathbb R}^{2d}_\theta$) and with some success: the dimensionally reduced Bogomolny equations are, for appropriate choice of parameters, integrable \cite{Lechtenfeld:2003cq}.  So it is natural to ask: what happens when the coset space $X$, instead of the physical space ${\cal M}$, is non-commutative, or both spaces are non-commutative? In particular, does the reduced theory still have vortices, and are they BPS? In this paper, we will focus on the case, where only the coset space $X$ is non-commutative.

A particular class of noncommutative coset spaces have been known for quite some time in the literature. Namely, these are the``fuzzy spaces'', of which the simplest and the most famous example is the fuzzy sphere, $S^2_F$ \cite{Madore, Book}. Gauge theory has been formulated on $S_F^2$ \cite{CarowWatamura:1998jn, Presnajder, Steinacker} and the group ${\rm SU(2)}$ acts naturally on it, so it seems well-suited for equivariant reduction.  Actually, equivariant reduction over fuzzy spaces has already been discussed in the literature, using the CSDR approach \cite{Aschieri:2003vy}.  However, only very simple examples have been studied so far, and not in great detail, so it seems important to try to perform an equivariant reduction in full.  In particular, one should compare equivariant reduction over fuzzy spaces with reduction over normal coset spaces to see what new features emerge.  It is worth mentioning that the fuzzy sphere appears in other gauge-theoretic contexts, such as the Aharony-Bergman-Jafferis-Maldacena model \cite{Nastase:2009ny}.  Equivariant reduction might prove a useful tool for constructing solutions to such models, perhaps along the lines of \cite{Arai:2008kv}.

With these motivations in mind, in this paper we present the fuzzy generalisation of Witten's equivariant reduction over ${\cal M} \times S^2$. To this end, we start from a ${\rm U(2)}$ Yang-Mills theory on ${\cal M} \times S_F^2$ and using essentially the representation theory of ${\rm SU(2)}$ we determine the most general ${\rm SU(2)}$-equivariant gauge field on ${\cal M} \times S_F^2$. This allows us to compute the reduced action in full. The latter appears to be an abelian Higgs-type model over  ${\cal M}$.  Specializing to a concrete and a simple case by selecting  ${\cal M} = {\mathbb R}^2$, we demonstrate that this model admits classical vortices and present their numerical solutions.

An outline of the rest of this paper is as follows: in section \ref{sec2} we will review gauge theory on ${\cal M}\times S^2_F$, in particular emphasising the approach in which it can be dynamically generated by a gauge theory on ${\cal M}$ with a larger gauge group.  In section \ref{sec3} we will review equivariant reduction over the fuzzy sphere, and give an explicit parametrisation of the equivariant gauge fields. In section \ref{sec4} we will carry out the reduction procedure, and give the reduced action explicitly. Section \ref{sec5} collects our analysis on the vacuum structure of the reduced theory, and section \ref{sec6} collects our results on its vortex solutions.  We summarise and comment on our results and mention some directions for future work in section \ref{sec7}.

\section{Yang-Mills Theory on ${\cal M} \times S_F^2$}
\label{sec2}

In this section, we collect the essential features of gauge theory on 
${\cal M} \times S_F^2$.  Actually, pure Yang-Mills theory on this space 
naturally appears as an effective description of a particular gauge theory 
with scalars on ${\cal M}$, as was recently pointed out in \cite{Aschieri:2006uw}.

We start by defining a gauge theory on ${\cal M}$.  Let $y^\mu$ be coordinates on 
${\cal M}$, let $A_\mu$ be $su({\cal N})$ valued anti-Hermitian
gauge fields and let $\phi_a \,, (a=1,2,3)$ be $3$ anti-Hermitian scalar
fields transforming in the adjoint of ${\rm SU(}{\cal N}{\rm )}$.  We introduce an action,
\begin{gather}
S = \int_{{\cal M} } d^d y  \, {\mbox Tr}_{{\cal N}} \Big (
\frac{1}{4g^2} F_{\mu \nu}^\dagger F_{\mu \nu} + 
(D_\mu \phi_a)^\dagger (D_\mu \phi_a) \Big ) + \frac{1}{\tilde{g}^2}V_1(\phi) +  a^2 V_2(\phi) \,,  \\
V_1(\phi) = \, {\mbox Tr}_{{\cal N}} \big ( F_{ab}^\dagger F_{ab} \big ) \, , \quad  
V_2(\phi) = \, {\mbox Tr}_{{\cal N}} \big ( (\phi_a \phi_a + {\tilde b})^2 \big ) 
\label{eq:actionfirst}
\end{gather}
Here $a$, $\tilde{b}$, $g$ and $\tilde{g}$ are constants and 
${\mbox Tr}_{{\cal N}} = {\cal N}^{-1} \mbox Tr$ denotes a normalised trace.  
In $V(\phi)$ we have used the definition
\be
F_{ab} := \lbrack \phi_a \,, \phi_b \rbrack - \varepsilon_{abc} \phi_c \,, 
\label{eq:curvaturefuzzy}
\ee
whose purpose will become evident shortly.

It is useful to note that $\phi_a$ transform in the vector
representation of an additional global $SO(3)$ symmetry, and that $V_1$ and $V_2$ are 
invariant under this symmetry.

This theory spontaneously develops extra dimensions in the form of fuzzy
spheres as formulated in detail in \cite{Aschieri:2006uw}. Let us very
briefly see how this actually comes about. We observe that the potential 
$\tilde{g}^{-2}V_1 +  a^2 V_2$ is positive definite, and that solutions of
\be
F_{ab} = \lbrack \phi_a \,, \phi_b \rbrack - \varepsilon_{abc} \phi_c
= 0 \,, \quad - \phi_a \phi_a = {\tilde b}
\label{eq:minimum1}
\ee
are evidently a global minima. A solution to these equations may be obtained by taking the value of
${\tilde b}$ as the quadratic Casimir of an irreducible representation of ${\rm SU(2)}$
labeled by $\ell$: ${\tilde b} = \ell (\ell + 1)$ with $2\ell\in\mathbb{Z}$. If we further
assume that the dimension ${\cal N}$ of the matrices $\phi_a$ 
is $(2 \ell +1) n$, then (\ref{eq:minimum1}) is solved by the
configurations of the form 
\be
\phi_a = X_a^{(2 \ell + 1)} \otimes {\bf 1}_n \,,
\label{eq:minimumsol}
\ee
where $X_a^{(2 \ell + 1)}$ are the (anti-Hermitian) generators of
${\rm SU(2)}$ in the irreducible representation $\ell$, which has dimension $2\ell+1$.
Here we have implicitly used the isomorphism $u((2\ell+1)n)\cong u(2\ell+1)\otimes u(n)$.  
We observe that this vacuum configuration spontaneously breaks the ${\rm SU}({\cal N})$ 
down to ${\rm U}(n)$ which is the commutant of $\phi_a$ in (\ref{eq:minimumsol}).
Fluctuations about this vacuum are described by a gauge theory on ${\cal M} \times S_F^2$, 
as we shall shortly see. 

We also wish to note that the most general solution to the equations
in (\ref{eq:minimum1}) is not known. However, a large class of solutions to these equations exist. They
are given by the block diagonal matrices
\be
\phi_a = diag \big ( \alpha_1 {(2 \ell_1 + 1)} \otimes {\bf 1}_{n_1} \,, \cdots \,, 
 \alpha_k {(2 \ell_k + 1)} \otimes {\bf 1}_{n_k} \big ) \,, 
\ee
such that ${\cal N} = n_1 (2 \ell_1 + 1) + \cdots + n_k (2 \ell_k + 1)$ and for some suitably chosen
constants $\alpha_i$. For instance, for $k=2$, this vacuum configuration leads to spontaneous breaking
of ${\rm SU}({\cal N})$ down to ${\rm SU}(n_1) \times {\rm SU}(n_2) \times {\rm U}(1)$. It turns out that
for $k=1$ and $k=2$, ${\rm SU}(n)$ and ${\rm SU}(n_1) \times {\rm SU}(n_2) \times {\rm U}(1)$
are the effective low-energy gauge groups of the reduced theories on ${\cal M}$, respectively.

For details of these results, and a discussion on another type of solution to the equations in (\ref{eq:minimum1})
with off-diagonal corrections, we refer the reader to the original article in \cite{Aschieri:2006uw}.  
Hereafter we will focus our attention on the vacuum configuration given in
(\ref{eq:minimumsol}).

The fuzzy sphere at level $\ell$ is defined to be the algebra of $(2\ell +1) \times (2 \ell +1)$ matrices 
$Mat(2 \ell +1)$.  The three Hermitian ``coordinate functions''
\be
{\hat x}_a := \frac{i}{\sqrt{\ell (\ell +1)}} X_a^{(2\ell+1)}
\ee
satisfy
\be
\lbrack {\hat x}_a \,, {\hat x}_b \rbrack = \frac{i}{\sqrt{\ell (\ell +1)}} \varepsilon_{abc} {\hat x}_c \,, 
\quad {\hat x}_a {\hat x}_a = 1 \,,
\ee
and generate the full matrix algebra $Mat(2 \ell +1)$.  There are three natural derivations of functions, defined by
the adjoint action of $su(2)$ on $S_F^2$:
\be
f \rightarrow ad X_a^{(2 \ell + 1)} f := \lbrack X_a^{(2 \ell + 1)} \,, f \rbrack \,, \quad f \in Mat(2 \ell +1) \,.  
\ee
In the limit $\ell\rightarrow\infty$, the functions $\hat{x}_a$ are identified with the standard coordinates 
$x_a$ on $\mathbb{R}^3$, restricted to the unit sphere and the infinite-dimensional algebra ${\cal C}^\infty(S^2)$ 
of functions on the sphere is recovered. Also in this limit, the derivations $[X_a^{(2 \ell + 1)},\cdot]$ become the vector 
fields $-i{\cal L}_a = \varepsilon_{abc} x_a \partial_b$ induced by the usual action of $SO(3)$.

Fluctuations about the vacuum (\ref{eq:minimumsol}) may be written
\be
\phi_a = X_a + A_a \,, 
\label{eq:config1} 
\ee
where $A_a \in u(2\ell+1)\otimes u(n)$ and we have abbreviated $X_a^{(2 \ell + 1)} \otimes {\bf 1}_n =: X_a$.  
Then $A_a$, $a=1,2,3$, may be interpreted as three components of a ${\rm U}(n)$ gauge field on the fuzzy sphere.
Thus, $\phi_a$ are the ``covariant coordinates'' on $S_F^2$ and (\ref{eq:curvaturefuzzy}) defines the associated
curvature $F_{ab}$. The latter may be expressed in terms of the gauge fields $A_a$ as:
\be
F_{ab} = \lbrack X_a \,, A_b \rbrack - \lbrack X_b \,, A_a \rbrack + \lbrack A_a \,, A_b \rbrack 
- \varepsilon_{abc} A_c \,.
\ee

The term $V_1$ is the obvious analog on the fuzzy sphere of the Yang-Mills action 
on the sphere.  However, with this term alone, gauge theory on the sphere is not recovered in the commutative limit, 
since the fuzzy gauge field has three components rather than two.  Rather, one obtains gauge theory with an additional scalar; the scalar is more precisely the component of the gauge field pointing in the radial direction when $S^2$ is embedded in $\mathbb{R}^3$.

The purpose of the term $V_2$ in the action is to suppress this scalar.  
To see how this works, observe that
\begin{multline}
i (\ell(\ell+1))^{-1/2} \Big( (X_a+A_a)(X_a+A_a)+\ell(\ell+1) \Big) = \{ \hat{x}_a, A_a \} + i (\ell(\ell+1))^{-1/2}A_a^2 \\
\xrightarrow[ \ell\rightarrow\infty]{}  2x_a A_a \,.
\end{multline}
The term $x_aA_a$ is precisely the component of the gauge field on the sphere associated with the radial direction, so the term 
$a^2V_2$ gives a mass $a\sqrt{\ell(\ell+1)}$ to this component.

It is possible to understand the origin of this mass term from the results of \cite{Aschieri:2006uw} in a non-trivial manner.
In the expansion of the scalar fields $\phi_a$ into modes, there is a mode corresponding to the fluctuations of the radius of $S_F^2$.
This is in fact the Higgs which acquires a positive mass after the spontaneous breaking of ${\rm SU}({\cal N})$ to ${\rm SU}(n)$. 
From the $V_2$ term in the potential this mass is determined to be $a\sqrt{\ell(\ell+1)}$, which is consistent with the predictions 
obtained from the $\ell \rightarrow \infty $ limit above.

To summarise, with (\ref{eq:config1}) the action in (\ref{eq:actionfirst}) takes the form of a ${\rm U(n)}$ gauge 
theory on ${\cal M} \times S_F^2(2 \ell + 1)$ with the gauge field components $A_M(y) = (A_\mu(y) \,, A_a(y))\in u(n)\otimes u(2\ell+1)$ and field strength tensor
\begin{eqnarray}
F_{\mu\nu} &=& \partial_\mu A_\nu - \partial_\nu A_\mu + [A_\mu,A_\nu]  \nn \\
F_{\mu a} &=& D_\mu \phi_a = \partial_\mu \phi_a + [A_\mu, \phi_a ] \\
F_{ab} &=& [\phi_a,\phi_b] - \epsilon_{abc}\phi_c \nn \,.
\end{eqnarray}
It is important to note that this gauge theory can only be considered ``standard'' Yang-Mills theory when the coefficients $g,\tilde{g}$ satisfy $g\tilde{g}=1$, for it is only in this case that the action takes the form of an $L^2$ norm of $F_{MN}$.  It is worth mentioning that even abelian gauge theory on the fuzzy sphere (the case $n=1$) is described by the non-abelian action (\ref{eq:actionfirst}), as was emphasised in \cite{Aschieri:2003vy}.

For future use we note that,
\be
{\mbox Tr}_{{\cal N}} = \frac{1}{n (2 \ell +1)} {\mbox Tr}_{Mat(2 \ell +1)} \otimes {\mbox Tr}_{Mat(n)}
\ee
where $Mat(k)$ denotes the algebra of $k \times k$ matrices.

In the following section we will focus on the case of a ${\rm U(2)}$ gauge theory on ${\cal M}\times S_F^2$, and 
explicitly construct the most general ${\rm SU(2)}$-equivariant gauge field on  $S_F^2$ using essentially 
the representation theory of ${\rm SU(2)}$. Subsequently, this will allows us to dimensionally reduce the 
gauge theory on ${\cal M} \times S_F^2$ to a ${\rm U(1)}$ abelian Higgs type model on ${\cal M}$.
We find that the latter may deviate from an abelian Higgs model on ${\cal M}$ which 
descends from dimensionally reducing the Yang-Mills theory on ${\cal M} \times S^2$.

\section{Finding the ${\rm SU(2)}$-Equivariant Gauge Field}
\label{sec3}

Equivariant dimensional reduction of gauge theories on coset spaces $G/H$ was
first formulated by Forgacs and Manton \cite{Forgacs}, see \cite{Zoupanos} for a review. 
The group $G$ acts naturally on the manifold ${\cal M}\times G/H$; the basic idea 
of Forgacs and Manton is to require that gauge fields are invariant under this 
action, up to a gauge transformation.  In this way, a gauge theory on ${\cal M}\times G/H$ 
can be reduced to a gauge theory on ${\cal M}$.  Their
treatment formalized an earlier result obtained by Witten
\cite{Witten}, whereby Yang-Mills theory on $\mathbb{R}^4$ was reduced to an abelian 
Higgs model on 2-dimensional hyperbolic space.

In recent times a general prescription for equivariant
reduction of gauge fields on ${\cal M} \times S^2_F$ has been described in
\cite{Aschieri:2003vy}, \cite{Aschieri:2006uw}.  In this article, we shall follow these 
articles' formalism, but choose a different action of the group ${\rm SU(2)}$.  We shall see 
later that our example reduces to Witten's ansatz in the commutative limit.  In this section, 
we shall outline the equivariant reduction formalism, and determine the most general 
${\rm SU(2)}$-equivariant gauge field on ${\cal M} \times S^2_F$ under our chosen action of ${\rm SU(2)}$.  

In all its generality, to carry out the ${\rm SU(2)}$-equivariant reduction
scheme, one chooses three elements $\omega_a\in u(2)\otimes u(2\ell+1)$ (for $a=1,2,3$),
and imposes the following symmetry constraints,
\be
\lbrack \omega_a \,, A_\mu \rbrack = 0  \,,
\label{eq:A}
\ee
\be
\lbrack \omega_a, \varphi_b \rbrack = \epsilon_{abc} \varphi_c,
\label{eq:vector}
\ee
on the gauge field. These constraints are consistent only if
$\omega_a$ satisfies:
\be
\lbrack \omega_a, \omega_b \rbrack = \varepsilon_{abc} \omega_c.
\ee

Apart from this restriction, we are free to select $\omega_a$
arbitrarily. In what follows, we shall choose
\be
\omega_a = X_a^{(2\ell+1)} \otimes {\bf 1}_2 - {\bf 1}_{2 \ell +1} \otimes
\frac{i\sigma^a}{2} \,.
\ee
These $\omega_a$ are the generators of the representation
$\underline{1/2} \otimes \underline{\ell}$ of ${\rm SU(2)}$, where by
$\underline{m}$ we denote the spin $m$ representation of ${\rm SU(2)}$, of
dimension $2m+1$.  The two terms which make up $\omega_a$ generate rotations and gauge transformations,
so imposing $\omega$-equivariance amounts to requiring that rotations can be compensated by gauge transformations. 
There are certainly more possible choices for $\omega_a$; for example $\omega_a=X_a^{(2\ell+1)} \otimes {\bf 1}_2$ was studied
in \cite{Aschieri:2003vy, Aschieri:2006uw}.

In order to study the dynamics of gauge fields subject to the
constraints (\ref{eq:A}), (\ref{eq:vector}), we shall first find a way
to parametrise their most general solution.  Once found, this
parametrisation will be substituted into the Yang-Mills action 
and by tracing over $S_F^2$ a reduced action on ${\cal M}$  
will be obtained. We also note that, by the principle of symmetry criticality \cite{Manton-Sutcliffe},
the equations of motion obtained from the reduced action will be the
same as the equations of motion that would have obtained by
substituting the parametrisation into the equations of motion of the
original Yang-Mills action.

Therefore, we will construct the most general solution of the symmetry
constraints, beginning with (\ref{eq:A}). The left hand side of this
equation tells us that $A_\mu$ transforms under the adjoint action of
$\omega_a$, or equivalently, in the representation $(\underline{1/2}
\otimes \underline{\ell}) \otimes (\underline{1/2} \otimes
\underline{\ell})$ of $su(2)$. The right hand side tells us that
$A_\mu$ belongs to a trivial sub-representation of this representation.
It is a simple application of the Clebsch-Gordan formula to find the
trivial sub-representations: for $\ell>1/2$, we find
\begin{eqnarray}
(\underline{1/2} \otimes \underline{\ell}) &\otimes& (\underline{1/2}
\otimes \underline{\ell}) \nn \\
&=& (\underline{\ell+1/2} \oplus \underline{\ell-1/2}) \otimes
(\underline{\ell+1/2} \oplus \underline{\ell-1/2}) \\
&=& (\underline{\ell+1/2}\otimes\underline{\ell+1/2}) \oplus
2(\underline{\ell+1/2}\otimes\underline{\ell-1/2}) \oplus
(\underline{\ell-1/2}\otimes\underline{\ell-1/2}) \nn \\
&=& 2\,\underline{0} \oplus 4\,\underline{1} \oplus\dots \nn 
\end{eqnarray}
Thus, the set of solutions to (\ref{eq:A}) is $2$-dimensional and a
convenient parametrisation is
\be
A_\mu = \frac{1}{2}Q a_\mu(y) + \frac{1}{2}i b_\mu(y)
\label{eq:amu}
\ee
In (\ref{eq:amu}) we have introduced the Hermitian ${\rm U(1)}$ gauge fields on ${\cal M}$:
\be
a_\mu^\dagger = a_\mu \,, \quad  b_\mu^\dagger = b_\mu \,,
\ee
and the anti-Hermitian, ``idempotent''\footnote{To be more accurate the idempotents are 
evidently $ \pm i Q$.} $Q$:
\begin{equation}
Q := \frac{X_a\otimes\sigma^a - i/2}{\ell+1/2} \,, \quad Q^\dagger = - Q \,,
\quad Q^2 = - {\bf 1}_{2(2 \ell +1)} \,.
\end{equation}
Indeed, $Q$ is the fuzzy version of $q := i {\bf \sigma} \cdot {\bf x}$ and
converges to it in the $\ell \rightarrow \infty$ limit. $\pm i Q$ appears
also in the context of monopoles and fermions over $S_F^2$ where 
in the former it is the idempotent associated with the projector describing
the rank $1$ monopole bundle over $S_F^2$, while in the latter it serves as the
chirality operator associated with the Dirac operator on $S_F^2$. For
further details on these topics we refer to the literature \cite{Book, Grosse, Bal}. 
  
We now proceed similarly with the constraint (\ref{eq:vector}). This
equation tells us that the vector $\phi_a$ belongs to a
$\underline{1}$ sub-representation of the representation
$(\underline{1/2} \otimes \underline{\ell}) \otimes (\underline{1/2}
\otimes \underline{\ell})$. Our calculation above shows that the
space of solutions has dimension $4$; an explicit parametrisation is
\begin{gather}
\phi_a = X_a + A_a \,, \nn \\
A_a = \frac{1}{2}\varphi_1(y)[X_a,Q] + \frac{1}{2} (\varphi_2(y)-1) Q[X_a,Q] 
+ i \frac{1}{2} \varphi_3(y) \frac{1}{2} \{ \hat{X}_a, Q \} +
\frac{1}{2} \varphi_4(y) \hat{\omega}_a.
\label{eq:eqvansatz}
\end{gather}
Here $\varphi_i$ are real scalar fields over ${\cal M}$, the curly brackets denote anti-commutators throughout, and we have further introduced
\be
\hat{X}_a := \frac{1}{\ell+1/2} X_a \,, \quad {\hat \omega}_a := \frac{1}{\ell+1/2} \omega_a.
\ee

It is worthwhile to remark that, in the commutative limit, (\ref{eq:eqvansatz}) becomes
\be
A_a \xrightarrow[\ell \rightarrow \infty]{} i \frac{1}{2}\varphi_1(y){\cal L}_a q  + i \frac{1}{2} (\varphi_2(y)-1) q {\cal L}_a q 
+ \frac{1}{2} \varphi_3(y) x_a q + \frac{1}{2} \varphi_4(y) x_a \,.
\label{ansatz}
\ee
In this limit, the component of $A_a$ normal to $S^2$ can be killed by imposing the constraint 
$x_a A_a = 0$ on the gauge field.  This constraint is satisfied if and only if we take $\varphi_3 =0 \,, \varphi_4=0$, 
as is easily observed from the above expression. Thus, we recover then the well-known expression for the 
spherically symmetric gauge field over ${\cal M} \times S^2$ \cite{Witten, Forgacs}.

\section{Dimensional Reduction of the Yang-Mills Action}
\label{sec4}

We are now in a position to substitute the ${\rm SU(2)}$-equivariant gauge field determined in the previous
section into the Yang-Mills action of section 2 and then trace over the fuzzy sphere to reduce 
it to an action on ${\cal M}$. It is quite important to note the following identities
\begin{gather}
\label{identity1}
\lbrace Q \,, \lbrack X_a \,, Q \rbrack \rbrace = 0 \,, \quad  \lbrace X_a \,, \lbrack X_a \,, Q \rbrack \rbrace = 0 \,,
\quad (\mbox{sum over repeated $a$ is implied}) \,, \\
\label{identity2}
\lbrack Q \,, \lbrace X_a \,, Q \rbrace \rbrack = 0 \,, \quad  \lbrack X_a \,, \lbrace X_a \,, Q \rbrace \rbrack = 0 \,,
\quad (\mbox{sum over repeated $a$ is implied}) \,.
\end{gather}
which significantly simplify the calculations, since they greatly reduce the number of traces to be computed.

The reduced action has the form
\be
S = \int_{\cal M} d^d y \,  {\cal L}_F + {\cal L}_G +  \frac{1}{\tilde{g}^2} V_1 +  a^2 V_2
\ee
These terms will be defined and explicitly evaluated below.

\subsection{The Field Strength Term}

The curvature term $F_{\mu \nu}$ associated with the connection $A_\mu$ takes the form
\begin{gather}
F_{\mu\nu} = \frac{1}{2} \left( f_{\mu\nu} Q + i h_{\mu\nu}  \right) \,, \nn \\
f_{\mu\nu}=\partial_\mu a_\nu-\partial_\nu a_\mu \,, \quad h_{\mu\nu}=\partial_\mu b_\nu-\partial_\nu b_\mu \,, \\
f_{\mu \nu}^\dagger = f_{\mu \nu} \,, \quad h_{\mu \nu}^\dagger = h_{\mu \nu} \nn \,.
\end{gather}
We find
\begin{eqnarray}
{\cal L}_F &:=& \frac{1}{4 g^2} {\mbox Tr}_{{\cal N}} \Big ( F_{\mu \nu}^\dagger F_{\mu \nu} \Big ) \nn \\
&=& \frac{1}{16 g^2} \Big (  f_{\mu\nu} f^{\mu\nu} +  h_{\mu\nu} h^{\mu\nu} + \frac{1}{\ell + \frac{1}{2}} f_{\mu\nu} h^{\mu\nu} \Big ) \,.
\end{eqnarray}

\subsection{The Gradient Term}

An easy calculation shows that
\be
D_{\mu}\phi_a = \frac{1}{2}(D_\mu \varphi_1 + Q D_\mu
\varphi_2)[X_a,Q] + \frac{i}{4}\partial_{\mu}\phi_3 \{ \hat{X}_a, Q \}
+ \frac{1}{2} \partial_\mu \varphi_4 \hat{\omega}_a
\ee
where we have used $D_\mu \varphi_i = \partial_\mu \varphi_i + \varepsilon_{ji} a_\mu \varphi_j$.  This formula demonstrates why the choice of parametrisation (\ref{ansatz}) is a good one: the identities (\ref{identity1}), (\ref{identity2}) imply that $\varphi_1+i\varphi_2$ is a complex scalar belonging to the fundamental representation of the gauge group ${\rm U(1)}$, while $\varphi_3$ and $\varphi_4$ a real scalars belonging to the trivial representation.

The gradient term in the action is then
\begin{multline}
{\cal L}_G := {\mbox Tr}_{{\cal N}} \Big ( (D_\mu \phi_a)^\dagger (D_\mu \phi_a) \Big ) \\
= \frac{1}{2} \frac{\ell^2+\ell}{(\ell+1/2)^2} \left( (D_\mu \varphi_1)^2 + (D_\mu
\varphi_2)^2 \right) 
+ \frac{1}{4} \frac{(\ell^2+\ell)(\ell^2+\ell-1/4))}{(\ell+1/2)^4}(\partial_\mu\varphi_3)^2 \\
+ \frac{1}{2}\frac{\ell^2+\ell}{(\ell+1/2)^3} \partial_\mu \varphi_3 \partial_\mu \varphi_4 
+ \frac{1}{4}\frac{\ell^2+\ell+3/4}{(\ell+1/2)^2} (\partial_\mu\varphi_4)^2.
\end{multline}

\subsection{The Potential Term}

It is easier to work with dual of the curvature $F_{ab}$ given by
\begin{multline}
\frac{1}{2} \varepsilon_{abc} F_{ab} = \frac{1}{2} \epsilon_{abc} \lbrack \phi_a,\phi_b \rbrack -\phi_c \,, \\
= \frac{1}{2}P_1 (\varphi_1+\varphi_2 Q)[X_c,Q] + \frac{i}{4} (|\varphi|^2-P_2) \frac{ \{X_c,Q\} }{(\ell+1/2)} 
+ \frac{1}{4}P_3 \frac{\omega_c}{(\ell+1/2)^2} \,.
\label{eq:dualcurvature}
\end{multline}
where $|\varphi|^2=\varphi_1^2+\varphi_2^2$, and $P_{1,2,3}$ are given in the appendix.

The potential term in the action may then be expressed as
\begin{equation}
V_1 =  \Big ( Q_1|\varphi|^4 + Q_2|\varphi|^2 + Q_3 \Big ) \,.
\end{equation}
The explicit expressions for $Q_{1,2,3}$ are given in the appendix.

In the large $\ell$ limit, we find   
\be
V_1 \underset{\ell \rightarrow \infty}{=}
\frac{1}{2}(|\varphi|^2+\varphi_3-1)^2 + \varphi_3^2|\varphi|^2 +
\frac{1}{2}\varphi_4^2 \,.
\ee

\subsection{The Constraint Term}

Firstly, following the discussion around (\ref{eq:minimum1}), we choose
${\tilde b} = \ell (\ell +1)$. With this input we can write
\begin{equation}
\label{constraint}
\phi_a \phi_a + \ell(\ell+1) = R_1+R_2iQ,
\end{equation}
where $R_1$ and $R_2$ are given in the appendix.

The constraint term in the action therefore takes the form 
\begin{eqnarray}
\label{constraint term}
V_2 &=& \Big ( R_1^2+R_2^2+\frac{1}{(\ell + \frac{1}{2})}R_1 R_2 \Big ).
\end{eqnarray}

\section{Vacua and topology}
\label{sec5}

In order to obtain a better understanding of the reduced action found in the previous section, we will analyse its vacua.  The potential has two parts:
\[ \frac{1}{\tilde{g}^2} V_1 + a^2 V_2. \]
Except in the case $a=0$, any zero of the potential must be a zero of both $V_1$ and $V_2$.  In order for topological vortex solutions to exist, it is crucial that the set $\mathcal{V}$ of vacua is not simply connected; we will see below that this is the case for the present situation.

$V_1$ is the $L^2$ norm of the curvature (\ref{eq:curvaturefuzzy}), so zeros of $V_1$ coincide with zeros of the curvature.  It is more practical to find zeros of the quadratic curvature than of the quartic $V_1$; accordingly, we determine the zeros of $V_1$ by solving the equations,
\begin{eqnarray}
\label{vac1} 0 &=& |\varphi| \left( \frac{\ell^2+\ell-1/4}{(\ell+1/2)^2}\varphi_3 + \frac{1}{\ell+1/2}\varphi_4 \right) \\
\label{vac2} |\varphi|^2 &=& (1-\varphi_3)\left( 1 + \frac{\varphi_4}{\ell+1/2} - \frac{\varphi_3}{2(\ell+1/2)^2} \right) \\
\label{vac3} 0 &=& \frac{\ell^2+\ell}{(\ell+1/2)^2} \left( \varphi_3^2-2\varphi_3 \right) + \varphi_4^2 + 2\frac{\ell^2+\ell-1/4}{\ell+1/2} \varphi_4.
\end{eqnarray}
It is not difficult to solve these algebraic equations; their solution set is $\bigcup_{i=1}^5\mathcal{V}_i$, where
\begin{eqnarray}
\mathcal{V}_1 &=& \{ |\varphi|=1, \varphi_3=0, \varphi_4=0 \} \,, \nn \\
\mathcal{V}_2 &=& \left\{ |\varphi|=1, \varphi_3=2, \varphi_4=-2 \frac{\ell^2+\ell-1/4}{\ell+1/2} \right\} \,, \nn \\
\mathcal{V}_3 &=& \left\{ |\varphi|=0, \varphi_3=1, \varphi_4=\frac{1}{2(\ell+1/2)} \right\} \,, \\
\mathcal{V}_4 &=& \left\{ |\varphi|=0, \varphi_3=1, \varphi_4=-2\frac{\ell^2+\ell}{(\ell+1/2)} \right\}  \,, \nn \\
\mathcal{V}_5 &=& \left\{ |\varphi|=0, \varphi_3=1 \pm (\ell+1/2), \varphi_4= -\frac{\ell^2+\ell-1/4}{\ell+1/2} \pm\frac{1}{2} \right\} \nn \,.
\end{eqnarray}

As a check on our calculations, we have substituted these values of $\varphi_i$ into the ansatz (\ref{eq:eqvansatz}) to find the covariant derivative $\phi_a$, making use of the identity
\be
X_a = \frac{1}{2} Q[X_a,Q] - \frac{i}{4(\ell + \frac{1}{2})}\{X_a,Q\} + 
\left( 1 - \frac{1}{4(\ell + \frac{1}{2})^2} \right)\omega_a \,.
\ee
We have found:
\begin{eqnarray}
\varphi_i \in\mathcal{V}_1 && \phi_a = \exp(\alpha Q) X_a \exp(-\alpha Q) \,, \quad 0\leq\alpha<\pi \,, \nn \\
\varphi_i \in\mathcal{V}_2 && \phi_a = \exp(\alpha Q) (-i\sigma^a/2) \exp(-\alpha Q) \,, \quad 0\leq\alpha<\pi \,, \nn \\
\varphi_i \in\mathcal{V}_3 && \phi_a = \omega_a \,, \\
\varphi_i \in\mathcal{V}_4 && \phi_a = 0 \,, \nn \\
\varphi_i \in\mathcal{V}_5 && \phi_a = \frac{1}{2}\omega_a \pm \frac{1}{4} \left( i\{X_a,Q\} + \frac{\omega_a}{\ell+1/2} \right)  \,. \nn
\end{eqnarray}
We have checked that these $\phi_a$ solve $[\phi_a,\phi_b]-\epsilon_{abc}\phi_c =0$, as they should.  This is obvious in the first four cases, in the fifth case the calculation is tricky but can be performed with some care.

Having determined the zeros of $V_1$, it is straightforward to substitute them into $V_2$ and hence determine the full set of vacua.  We find that $V_2$ is zero only on the subset $\mathcal{V}_1$ of $\bigcup_{i=1}^5\mathcal{V}_i$, so the set of vacua is $\mathcal{V}=\mathcal{V}_1$.  In particular $\pi_1(\mathcal{V})=\mathbb{Z}$, so if ${\cal M}=\mathbb{R}^2$ for example, finite action configurations are classified by an integer-valued topological charge, the winding number of $\varphi_a : S^1_\infty\rightarrow \mathcal{V}$.

We remark here that, while we have shown that ${\rm SU(2)}$-equivariant instantons on $\mathbb{R}^2\times S^2_F$ are classified by a single integer topological charge, there is no reason to expect that the same holds for non-equivariant instantons, even when $S^2_F$ is replaced by $S^2$.  Indeed, it seems quite likely that non-equivariant instantons on $\mathbb{R}^2\times S^2$ or $\mathbb{R}^2\times S^2_F$ have fractional charge, for the following reason.  In general, the topological charge of an instanton is equal to the Chern-Simons invariant of the connection induced on the manifold at infinity (which is usually flat).  Since the manifolds at infinity of $\mathbb{R}^2\times S^2$ and of $\mathbb{R}^3\times S^1$ are both $S^1\times S^2$, we expect instantons on both of these spaces to have similar topological classifications.  But instantons on $\mathbb{R}^3\times S^1$ can have non-integer charge \cite{gross}, therefore one expects the same to be true of instantons on $\mathbb{R}^2\times S^2$ or $\mathbb{R}^2\times S^2_F$.  However, we don't know of any example of an instanton with non-integer charge on these spaces.

\section{Vortices}
\label{sec6}

In this section we study vortex solutions to the Euler-Lagrange equations derived from the dimensionally-reduced action.  For simplicity, we restrict attention to the case ${\cal M}=\mathbb{R}^2$.  We ultimately restrict attention to ``standard'' Yang-Mills theory, with coupling constants $g=1/\sqrt{2}$, $\tilde{g}=\sqrt{2}$.  There is no canonical choice for the coefficient $a^2$ of the fuzzy constraint term; here we consider only the extreme cases of $a^2=0$ and $a^2=\infty$, which correspond respectively to imposing no constraint at all, and to imposing the constraint $\phi_a\phi_a+\ell(\ell+1)=0$ ``by hand''.  Finally, we assume that $\ell$ is large.  In the case $a=0$, we assume $\ell=\infty$ since this already constitutes a novel model.  In the $a=\infty$ theory we include only terms appearing at $O(\ell^{-2})$.

\subsection{Case 1: No constraint}

With $a=0$ and $\ell=\infty$, the action reduces to
\begin{multline}
S = \int_{\mathbb{R}^2} d^2y \, \frac{1}{16g^2} (f_{\mu\nu}f^{\mu\nu} + h_{\mu\nu}h^{\mu\nu} ) + \frac{1}{2} |D_\mu\varphi|^2 + \frac{1}{4}(\partial_\mu\varphi^3)^2 + \frac{1}{4}(\partial_\mu\varphi^4)^2 \\
+ \frac{1}{\tilde{g}^2} \left( \frac{1}{2} (|\varphi|^2+\varphi_3-1)^2 + |\varphi|^2\varphi_3^2 + \frac{1}{2}\varphi_4^2 \right) \,.
\end{multline}
The fields $\varphi_4$ and $b_\mu$ decouple from the rest, and may consistently be set to zero.

For the remaining fields we make the standard rotationally symmetric ansatz \cite{Manton-Sutcliffe} : we choose a gauge so that $a_r=0$ and set $\varphi = \chi(r) \exp(iN\theta)$, $\varphi_3=\lambda(r)$, $a = a_\theta(r) d\theta$, where $(y^1,y^2)=r(\cos\theta,\sin\theta)$.  The action reduces to,
\begin{multline}
S = 2\pi \int_0^\infty dr \frac{1}{8g^2r} a_\theta'^2 + \frac{r}{2} \chi'^2 + \frac{1}{2r} (N+a_\theta)^2\chi^2 + \frac{r}{4} \lambda'^2 \\
+ \frac{r}{\tilde{g}^2} \left( \frac{1}{2} (\chi^2+\lambda-1)^2 + \chi^2\lambda^2  \right) \,.
\end{multline}
The Euler-Lagrange equations obtained from this integral are
\begin{eqnarray}
\nonumber
0 &=& \chi'' + \frac{1}{r}\chi' -  \left( \frac{1}{r^2}(N+a_\theta)^2 + \frac{2}{\tilde{g}^2} (\chi^2+\lambda-1 + \lambda^2) \right)\chi \\
\label{ele1}
0 &=& a_\theta'' - \frac{1}{r}a_\theta' - 4g^2(a_\theta+N)\chi^2   \\
\nonumber
0 &=& \lambda'' + \frac{1}{r} \lambda' - \frac{2}{\tilde{g}^2} (\chi^2+\lambda-1+2\chi^2\lambda).
\end{eqnarray}

We have not found any analytic solutions to these equations.  However, as we shall see below, they are amenable to the usual approximation methods: one can obtain approximate solutions in the regions of small and large $r$, and one can solve the equations numerically.

\begin{table}[tb]
\begin{center}
\begin{tabular}{c|c|ccc|cc}
$N$ & $S/\pi$ & $\chi_0$ & $a_0$ & $\lambda_0$ & $C_1$ & $C_3$ \\
\hline
1 & 0.894 & 0.657 & 0.399 & 0.402 & 1.38 & 3.17 \\
2 & 1.618 & 0.212 & 0.330 & 0.666 & 5.53 & 14.5
\end{tabular}
\caption{The value of the action $S$ for vortices with $a=0$ and $N=1,2$, and constants associated with asymptotic expansions}
\label{table1}
\end{center}
\end{table}

\begin{figure}[tb]
\epsfig{file = 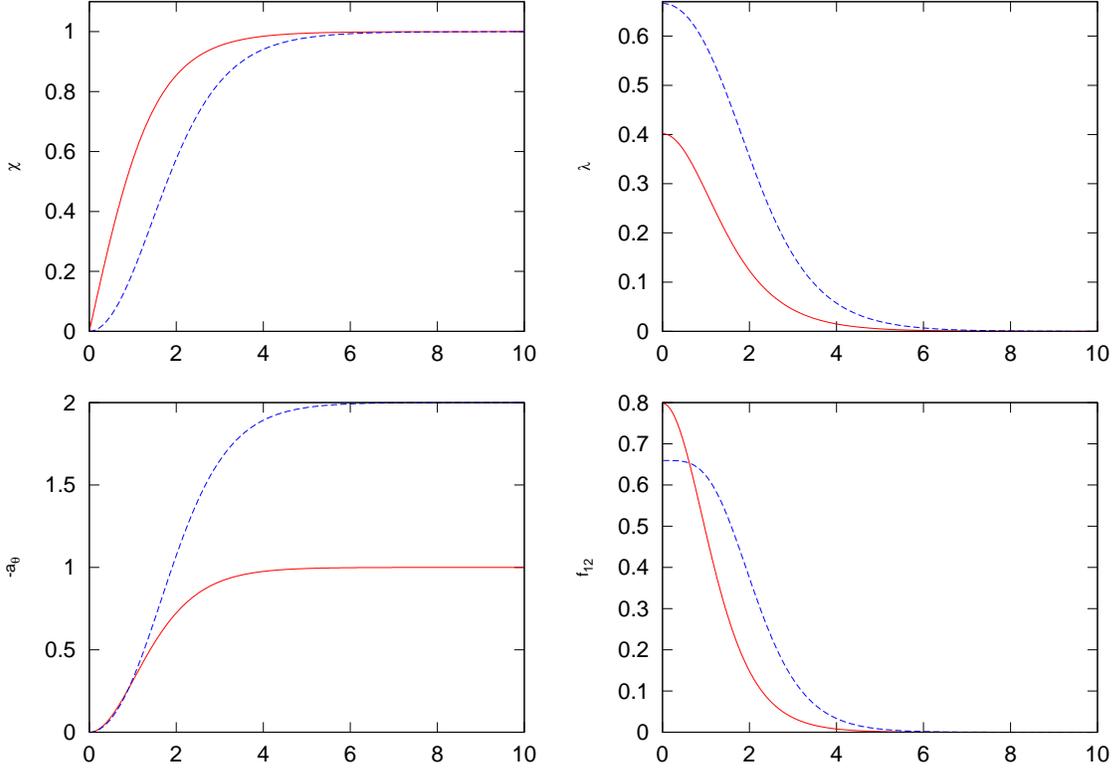, scale = 1.2}
\caption{The axially symmetric vortex with no constraint.  Graphs show $\chi$, $\lambda$, $-a_\theta$, and $f_{12}=\partial_ra_\theta/r$ as functions of r, for $N=1$ (solid) and $N=2$ (dashed).}
\label{figure1}
\end{figure}

Continuity of the fields implies that $\chi=O(r)$ and $a_\theta=O(r^2)$ as $r\rightarrow0$.  The equations (\ref{ele1}) imply further that around $r=0$ the following expansions hold for constants $\chi_0$, $a_0$, $\lambda_0$:
\begin{eqnarray}
\chi &=& \chi_0r^N + O(r^{N+2}) \,, \nn \\
a_\theta &=& a_0r^2 + O(r^4) \,, \\
\lambda &=& \lambda_0 + O(r^2) \,. \nn
\end{eqnarray}

Finiteness of the action implies that $\chi(r)\rightarrow1$, $a_\theta(r)\rightarrow-N$, and $\lambda(r)\rightarrow0$ as $r\rightarrow\infty$.  Accordingly, we set $\chi= 1-\delta\chi$, and $a_\theta=-N+\delta a$.  Under the assumption that $(\delta a/r)^2$ is subleading to $\delta\chi$ and $\lambda$, the Euler-Lagrange equations (\ref{ele1}) have the following large $r$ expansions:
\begin{eqnarray}
0 &=& \delta\chi'' + \frac{1}{r}\delta\chi' - \frac{2}{\tilde{g}^2}(-\lambda+2\delta\chi) \,, \nn \\
0 &=& \delta a'' - \frac{1}{r}\delta a' - 4g^2\delta a \,, \\
0 &=& \lambda'' + \frac{1}{r}\lambda' - \frac{2}{\tilde{g}^2}(3\lambda-2\delta\chi) \nn \,.
\end{eqnarray}
These equation can be solved in terms of modified Bessel functions $K_\alpha$ and constants $C_i$:
\begin{eqnarray}
\delta\chi &=& C_1 K_0 \left( \frac{\sqrt{2} r}{\tilde{g}} \right) - C_2 K_0 \left( \frac{2\sqrt{2} r}{\tilde{g}} \right) \,, \nn \\
\delta a &=& C_3 r K_1(2gr) \,, \\
\lambda &=& C_1 K_0 \left( \frac{\sqrt{2} r}{\tilde{g}} \right) + 2C_2 K_0 \left( \frac{2\sqrt{2} r}{\tilde{g}} \right) \,. \nn 
\end{eqnarray}
Of course, the terms with coefficient $C_2$ can be ignored at large $r$ since they are subleading.  Notice that our assumption that $(\delta a/r)^2$ is subleading to $\delta\chi$ and $\lambda$ is satisfied provided that $4g>\sqrt{2}/\tilde{g}$.  This holds for example when $g=1/\sqrt{2}$ and $\tilde{g}=\sqrt{2}$.  Notice also that the field strength decays faster than the scalars for these values of the coupling constants.  Since the field strength and scalars are respectively responsible for repulsive and attractive forces between vortices, this result indicates that vortices will attract in this model.

Finally, we present our numerical results.  We have solved the equations (\ref{ele1}) using the Runge-Kutta order 4 method.  The equations were studied on a finite interval of length $L$.  The expansions around $r=0$ were used as initial data, and the constants $\chi_0,a_0,\lambda_0$ are determined by the requirement that $\chi=1$, $a_\theta=-N$ and $\lambda=0$ at $r=L$.  We have computed the action $S$ of the resulting fields, as well as the coefficients of the asymptotic expansions, for a few values of $N$.  The results were independent of the length $L$ and the lattice spacing $h$, for sufficiently large $L$ and small $h$.  The constant $C_1$ was computed both from $\chi$ and $\lambda$, and the values obtained agreed.  Our results are summarised in table \ref{table1} and the numerical solutions are displayed in figure \ref{figure1}.

The main result of the numerical computation is that the value of the ratio $S/N$ is smaller for a symmetric $N=2$ vortex than for a symmetric $N=1$ vortex, suggesting again that vortices attract in this model.  It seems plausible that the symmetric vortex is the minimum amongst $N=2$ configurations, but this cannot be verified without further analysis.

We emphasise that the results in this section apply only to the case $\ell=\infty$.  An obvious next step would be to repeat this analysis look at the theory at $O(1/\ell)$.  We have written the $1/\ell$ correction to the action in the appendix; however, we haven't attempted to perform any numerical analysis on this theory, since we don't expect its behaviour to differ qualitatively from the $\ell=\infty$ case.

\subsection{Case 2: The constraint fully imposed}

We observe that the fuzzy constraint $\phi_a\phi_a+\ell(\ell+1)=0$ is equivalent to the two algebraic equations, $R_1=0$, $R_2=0$. These can be solved to obtain $\varphi_3$ and $\varphi_4$ in terms of $\varphi_1$ and $\varphi_2$.  Substituting back into the action yields an action with just one complex scalar field $\varphi=\varphi_1+i\varphi_2$.

When $\ell=\infty$, the solution to the constraint is simply $\varphi_3=0$, $\varphi_4=0$, and substituting these into the action yields the standard critically coupled Ginzburg-Landau energy functional.  When $\ell$ is large but finite, one can solve the constraint approximately by expanding about the $\ell=\infty$ solution in powers of $1/\ell$.  To leading order, this approximate solution is
\be
\varphi_3 = -\frac{1}{2\ell^2}(1-|\varphi|^2) + O\left(\frac{1}{\ell^3}\right) \,, \quad  \varphi_4 = \frac{1}{2\ell}(1-|\varphi|^2) + O\left(\frac{1}{\ell^2}\right)\,.
\ee
Taking now $g=1/\sqrt{2}$ and $\tilde{g}=\sqrt{2}$ and substituting the approximate solution above into the ansatz determines 
the leading order correction to the action:
\begin{multline}
S = \frac{1}{2} \int_{\mathbb{R}^2} d^2y \, \frac{1}{4}\left( f_{\mu\nu}f^{\mu\nu} + h_{\mu\nu} h^{\mu\nu} + \frac{1}{\ell} h_{\mu\nu} f^{\mu\nu} \right) + \left( 1 - \frac{1}{4\ell^2}\right) |D_\mu\varphi|^2 \\
+ \frac{1}{8\ell^2}(\partial_\mu|\varphi|^2)^2 + \frac{1}{2} \left(1+\frac{1}{2\ell^2}\right)(1-|\varphi|^2)^2 \,.
\end{multline}
The equation of motion for $h_{\mu\nu}$ is solved by $h_{\mu\nu}=-f_{\mu\nu}/(2\ell)$, and substituting back gives
\begin{multline}
S = \frac{1}{2} \int_{\mathbb{R}^2}  d^2y \, \frac{1}{4}\left( 1-\frac{1}{4\ell^2} \right) f_{\mu\nu}f^{\mu\nu} + \left( 1 - \frac{1}{4\ell^2}\right) |D_\mu\varphi|^2 \\
+ \frac{1}{8\ell^2}(\partial_\mu|\varphi|^2)^2 + \frac{1}{2} \left(1+\frac{1}{2\ell^2}\right)(1-|\varphi|^2)^2 \,.
\label{eq:conaction}
\end{multline}

With $\ell=\infty$, the standard Ginzburg-Landau energy functional is recovered, as is evident from (\ref{eq:conaction})
An interesting feature of the perturbed action is that the kinetic term for $\varphi=\varphi_1+i \varphi_2$ is non-linear.  This arises simply because the fields $\varphi_1,\varphi_2$ take values in the curved 2-manifold of solutions to the fuzzy constraint in $\mathbb{R}^4$.

\begin{table}[tb]
\begin{center}
\begin{tabular}{c|c|cc|cc}
$N$ & $S/\pi$ & $\chi_0$ & $a_0$ & $D_1$ & $D_2$ \\
\hline
1 & $1.00 - 0.0406\ell^{-2}$ & $0.853+0.284 \ell^{-2}$ & $0.500+0.128 \ell^{-2}$ & $1.71-0.68 \ell^{-2}$ & $2.42-0.48\ell^{-2}$ \\
2 & $2.00-0.0147 \ell^{-2}$ & $0.459+0.247 \ell^{-2}$ & $0.500+0.145 \ell^{-2}$ & $5.34-2.42 \ell^{-2}$ & $7.55-2.62 \ell^{-2}$
\end{tabular}
\caption{The value of the action $S$ for vortices with $a=\infty$ and $N=1,2$, and constants associated with asymptotic expansions}
\label{table2}
\end{center}
\end{table}

\begin{figure}[tb]
\epsfig{file = 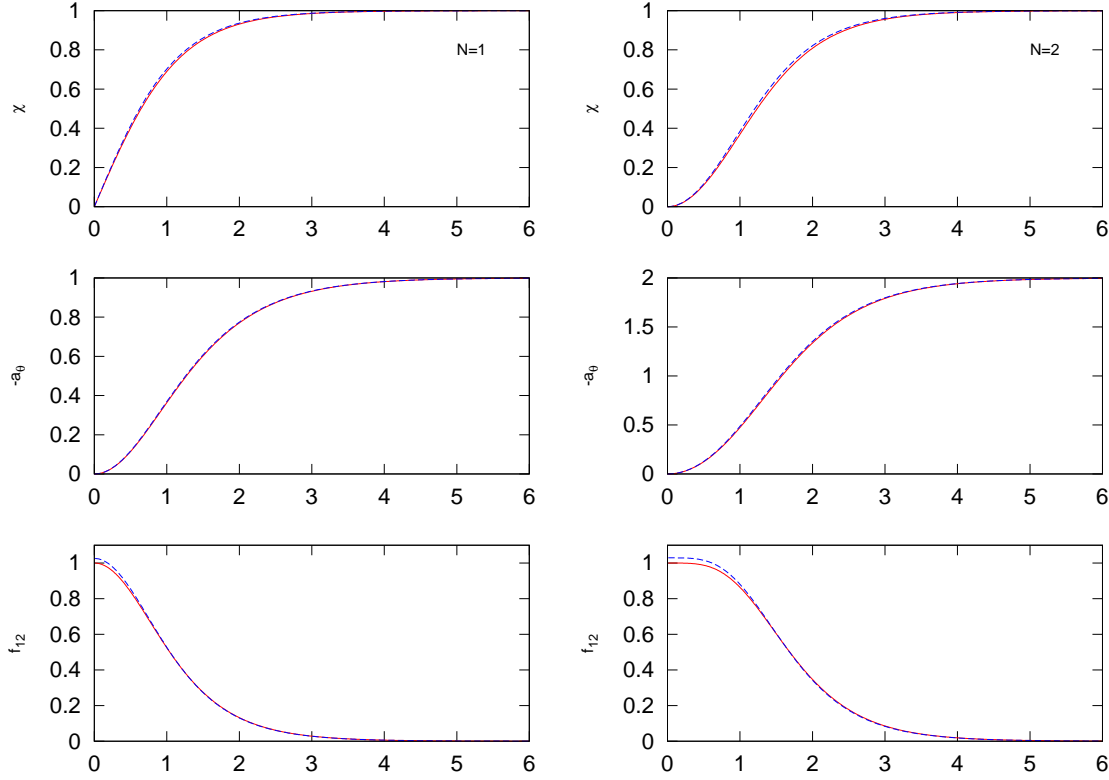, scale = 1.2}
\caption{The axially symmetric vortex with constraint fully imposed.  The left column corresponds to $N=1$, and the right to $N=2$.  Graphs show $\chi$, $-a_\theta$, and $f_{12}=\partial_ra_\theta/r$ as functions of r, for $\ell^{-2}=0$ (solid) and $\ell^{-2}=0.1$ (dashed).}
\label{figure2}
\end{figure}

In order to look for vortex solutions, we again make a radial ansatz $a_r=0$, $\varphi = \chi(r) \exp(iN\theta)$, $a = a_\theta(r) d\theta$.  Substituting into the action yields
\begin{multline}
S = \pi \int_0^\infty dr \, \left(1-\frac{1}{4\ell^2}\right) \left( \frac{1}{2r}a_\theta'^2 + r\chi'^2 + \frac{1}{r}(N+a_\theta)^2\chi^2 \right) \\
+ \frac{r}{2\ell^2} \chi^2\chi'^2 + \left(1+\frac{1}{2\ell^2}\right) \frac{r}{2} (1-\chi^2)^2 \,.
\end{multline}
The Euler-Lagrange equations for this functional are
\beqa
0 &=& \left(1 - \frac{1}{4\ell^2}+\frac{\chi^2}{2\ell^2}\right) \left(\chi'' + \frac{1}{r}\chi'\right) + \frac{1}{2\ell^2}\chi'^2\chi \nn  \\
&&- \left( \left(1-\frac{1}{4\ell^2}\right)\frac{1}{r^2}(N+a_\theta)^2 + \left(1+\frac{1}{2\ell^2}\right)(\chi^2-1) \right)\chi \,, \\
0 &=& a_\theta'' - \frac{1}{r}a_\theta' - 2(a_\theta+N)\chi^2 \,.
\eeqa

The perturbative solution about $r=0$ is
\begin{eqnarray}
\chi &=& \chi_0r^N + O(r^{N+2}) \,, \nn \\
a_\theta &=& a_0r^2 + O(r^4) \,. 
\end{eqnarray}

With $\chi=1-\delta\chi$, $a_\theta=-N+\delta a$, the large $r$ expansion of the Euler-Lagrange equations is
\begin{eqnarray}
0 &=& \left(1 + \frac{1}{4\ell^2}\right) \left(\chi'' + \frac{1}{r}\chi'\right) - 2\left(1+\frac{1}{2\ell^2}\right)\delta\chi \,,\nn \\
0 &=& \delta a'' - \frac{1}{r}\delta a' - 2\delta a \,.
\end{eqnarray}
These equations are solved by
\begin{eqnarray}
\delta\chi &=& D_1 K_0\left( \sqrt{2\frac{1+1/(2\ell^2)}{1+1/(4\ell^2)}}r \right)\,, \nn \\
\delta a &=& D_2rK_1(\sqrt{2}r) \,.
\end{eqnarray}
Notice that, for $1/\ell\neq0$, the scalar $\delta\chi$ decays faster than the field strength.  Since the field strength is responsible for a repulsive force, this indicates that vortices will repel in this model.

Finally, we have found numerical solutions to the Euler-Lagrange equations for a range of values of $1/\ell$, using the same method as in the previous subsection.  With $1/\ell=0$ our data agrees with established results \cite{Manton-Sutcliffe}; in table \ref{table2} we display the $1/\ell=0$ results, together with the leading order correction.  The numerical solutions are displayed in figure \ref{figure2}.

Notice that when $1/\ell\neq0$, the value of the ratio $S/N$ is larger for $N=2$ than for $N=1$, suggesting that the axially symmetric 2-vortex is unstable.  The simplest possible interpretation of our numerical and analytical results is that the axially symmetric 2-vortex is unstable to decay into two 1-vortices, which repel until they reach infinite separation.  However, more complicated behaviour is not ruled out -- for example, the axially symmetric 2-vortex could be a local minimum of the action, or it could decay into a stable non-symmetric configuration.

\section{Conclusion}
\label{sec7}

In this paper, following the fuzzy generalization of the CSDR scheme, we have first determined the most general ${\rm SU(2)}$ equivariant gauge connection over  ${\cal M} \times S^2_F$ and used it to dimensionally reduce the Yang-Mills theory over this space to an abelian Higgs-type theory over ${\cal M}$. Our results explicitly confirm that successful CSDR schemes can be implemented in the fuzzy setting. The main difference in the fuzzy scheme compared with standard CSDR is that additional degrees of freedom are present in the ${\rm SU(2)}$ equivariant gauge connection and they contribute as additional real scalars in the reduced theory. We have seen that, this new feature of the reduced theory can be successfully attributed to the fact that the gauge field on the fuzzy sphere has three components rather than two. These new real scalars appearing in the reduced action can be suppressed by including a constraint term in the Yang-Mills action, which gives a mass to one component of the gauge field.  If this mass is chosen very large, the reduced action obtained from the fuzzy reduction is very similar to the reduced action obtained from the standard CSDR.

We have also found analytical and numerical evidence for vortices in the reduced theory over $\mathbb{R}^2$, which map back to instantons on $\mathbb{R}^2\times S^2_F$.  However, the vortices obtained in the fuzzy reduction are not BPS, unlike in the standard reduction.  This fact can again be attributed to the gauge field on the fuzzy sphere having three, rather than two, components.  The self-dual equation for instantons is intrinsically 4-dimensional, so it doesn't make sense for a gauge field on $\mathbb{R}^2\times S^2_F$ with 5 components to be self-dual.  The fuzzy constraint, while removing one component of the gauge field, still doesn't seem to allow any BPS property.

Instead of being BPS, the vortices in the reduced model either attract or repel, according to whether the parameter $a$ is $0$ or $\infty$.  One might hope that for some intermediate value of $a$ critically coupled vortices exist; however, we doubt that this is the case, since we have not found a natural self-dual equation on $\mathbb{R}^2\times S^2_F$.  We believe that the vortices in the reduced theories deserve more study.  Apart from a more rigorous analysis of their stability and interactions, it would be interesting to see whether the additional scalars allow the existence of ``super-conducting strings'' \cite{Witten:1984eb}, or even more exotic solutions. 

Much of our analysis has focused on the case where $\ell$ is large, so it might prove fruitful to study the same reduction from a small $\ell$ point of view: for example, by fixing $\ell$ and working directly with matrices rather than algebraic identities.  It is possible that other interesting new features may emerge in this case. We  also would like to mention briefly that vortices have also recently been studied in the context of Yang-Mills theory on ${\cal M}\times X$, with $X$ chosen to be discrete 2-point space \cite{Otsu:2009vy}. More work is necessary to reveal  points of contact of this study with present developments, if there are any.

There are several other interesting questions which remain to be studied.  Recently, there has been some new developments in incorporating fermions into fuzzy reduction schemes \cite{Dolan-Szabo, Steinacker2} (see also \cite{Dolan:2007uf} for related developments), so it would be definitely interesting to try to incorporate the fermions into the example presented here. It would also be worthwhile to perform the the dimensional reduction on  ${\cal M}_ \times S_F^2$ where ${\cal M}$ too is a non-commutative manifold such as the $2d$-dimensional Moyal space ${\mathbb R}^{2d}_\theta$ and compare our results with those of the references \cite{Lechtenfeld:2003cq} in which the reduction over ${\mathbb R}^{2d}_\theta \times S^2$ is considered and non-commutative BPS vortices over  ${\mathbb R}^{2d}_\theta$ have been found. Progress on these topics will be reported elsewhere.

\vskip 1em

{\bf Acknowledgements}

\vskip 1em

We thank O. Lechtenfeld and A. D. Popov for useful comments and suggestions. S.K. was supported by the cluster of excellence EXC 201 {\it QUEST} of the Leibniz Universit\"at Hannover and by the Deutsche Forschungsgemeinschaft (DFG) under Grant No. LE 838/9.  D.H. is supported by the Graduiertenkolleg 1463 {\it Analysis, Geometry and String Theory}.

\appendices

\subsection{Explicit Formulae}

In this appendix, we list the explicit expressions for $P_1\,, P_2\,, P_3$, $Q_1\,, Q_2\,, Q_3$ and $R_1\,, R_2$
which were introduced for brevity of notation in section 5. 

We have
\beqa
P_1 &=& \frac{\ell^2+\ell-1/4}{(\ell+1/2)^2}\varphi_3 + \frac{1}{\ell+1/2}\varphi_4 \,, \\
P_2 &=& (1-\varphi_3)\left( 1 + \frac{\varphi_4}{\ell+1/2} - \frac{\varphi_3}{2(\ell+1/2)^2} \right) \,,\\
P_3 &=& \frac{\ell^2+\ell}{(\ell+1/2)^2} \left( \varphi_3^2-2\varphi_3
\right) + \varphi_4^2 + 2\frac{\ell^2+\ell-1/4}{\ell+1/2} \varphi_4 \,.
\eeqa

$Q_{1,2,3}$ are given in terms of the above $P_{1,2,3}$
\begin{eqnarray}
Q_1 &=& \frac{1}{2} \frac{(\ell^2+\ell)(\ell^2+\ell-1/4)}{(\ell+1/2)^4} \,, \\
Q_2 &=& \frac{\ell^2+\ell}{(\ell+1/2)^2} \left( P_1^2 -
  \frac{\ell^2+\ell-1/4}{(\ell+1/2)^2}P_2 + \frac{1}{2(\ell+1/2)^2}
  P_3 \right) \\
&=& -\frac{(\ell^2+\ell)(\ell^2+\ell-1/4)}{(\ell+1/2)^4} +
\frac{(\ell^2+\ell)(\ell^2+\ell-3/4)}{(\ell+1/2)^4} \varphi_3 \nn \\
&& + \frac{ (\ell^2+\ell) ((\ell+1/2)^4-(\ell+1/2)^2+3/8) }{(\ell+1/2)^6} \varphi_3^2 \nn \\
&& + 3 \frac{(\ell^2+\ell)(\ell^2+\ell-1/4)}{(\ell+1/2)^5}\varphi_3\varphi_4
+ \frac{3}{2}\frac{\ell^2+\ell}{(\ell+1/2)^4}\varphi_4^2 \,, \nn \\
Q_3 &=& \frac{1}{2}
\frac{(\ell^2+\ell)(\ell^2+\ell-1/4)}{(\ell+1/2)^4} P_2^2 +
\frac{1}{8} \frac{\ell^2+\ell+3/4}{(\ell+1/2)^4} P_3^2 -
\frac{1}{2}\frac{\ell^2+\ell}{(\ell+1/2)^4} P_2P_3 \,. 
\end{eqnarray}

For $R_1$ and $R_2$ we find
\begin{multline}
R_1 = -\frac{1}{2}(\varphi_1^2+\varphi_2^2-1) - \frac{1}{4(\ell +
  \frac{1}{2})^2}\varphi_3 - \left((\ell +
  \frac{1}{2})-\frac{1}{2(\ell + \frac{1}{2})}\right) \varphi_4 \\ 
- \left( \frac{1}{4} - \frac{3}{16(\ell + \frac{1}{2})^2} \right) \varphi_3^2
- \frac{1}{4(\ell + \frac{1}{2})}\varphi_3\varphi_4 - \frac{1}{4}\varphi_4^2  \,,
\end{multline}
\begin{multline}
R_2 = \frac{1}{4(\ell + \frac{1}{2})}(\varphi_1^2+\varphi_2^2-1) -
\left( (\ell + \frac{1}{2}) - \frac{3}{4(\ell + \frac{1}{2})}\right)
\varphi_3 - \frac{1}{2}\varphi_4 
- \frac{1}{16(\ell + \frac{1}{2})^3}\varphi_3^2 \\
- \left( \frac{1}{2} - \frac{1}{4(\ell + \frac{1}{2})^2}
\right)\varphi_3\varphi_4 - \frac{1}{4(\ell + \frac{1}{2})}\varphi_4^2 \,.
\end{multline}


\setcounter{equation}{0}

\subsection{Reduced Action at order $\frac{1}{\ell}$}

At order $\frac{1}{\ell}$ the reduced action takes the form
\begin{multline}
S = \int_{\cal M} d^d y \, \, 
\frac{1}{16 g^2} ( f_{\mu \nu} f^{\mu \nu} +  h_{\mu \nu} h^{\mu \nu} + \frac{1}{\ell} h_{\mu \nu} f^{\mu \nu} )  
+ \frac{1}{2} | D_\mu \varphi|^2  + \frac{1}{4} \Big(1 - \frac{1}{\ell} \Big)(\partial_\mu \varphi_3)^2 \\ 
+ \frac{1}{4} (\partial_\mu \varphi_4)^2  + \frac{1}{2 \ell} (\partial_\mu \varphi_3) (\partial_\mu \varphi_4) 
+ V_1 \Big|_{\frac{1}{\ell}}  + V_2 \Big|_{\frac{1}{\ell}}  + O \left (\frac{1}{\ell^2}\right)  \,.
\end{multline}
where

\begin{multline}
V_1 \Big|_{\frac{1}{\ell}} = \frac{1}{{\tilde g}^2} \big (
\frac{1}{2}(|\varphi|^2+\varphi_3-1)^2 + \varphi_3^2|\varphi|^2 + 
\frac{1}{2}\varphi_4^2 \big) \\
+ \frac{1}{\ell} \Big ( 3 |\varphi|^2 \varphi_3 \varphi_4 + \frac{1}{2} \varphi_4
(\varphi_3^2 - 2 \varphi_3 + \varphi_4) \Big ) \,. 
\end{multline}
and
\begin{multline}
V_2 \Big|_{\frac{1}{\ell}} = a^2 \Bigg ( (|\varphi|^2 - 1)^2 + \frac{1}{16} \varphi_3^4
+ \frac{1}{4} \varphi_3^3 + \big ( (\ell + \frac{1}{2})^2 - \frac{3}{2} \big) \varphi_3^2
+  \frac{1}{16} \varphi_4^4  
+ \frac{1}{2} (\ell + 1 + \frac{3}{4 \ell}) \varphi_4^3 \\
+ \ell (\ell + 1) \varphi_4^2 
+ (|\varphi|^2 - 1)^2 \big ( \frac{1}{4} (\varphi_3 ^2 + \varphi_4^2) + \frac{1}{4 \ell} \varphi_3 \varphi_4 +
(\ell + \frac{1}{2} - \frac{3}{4 \ell}) \varphi_4 \big ) 
+ \frac{1}{4 \ell} \varphi_3^3 \varphi_4\\
+ \frac{3}{8}\varphi_3^2 \varphi_4^2 + ( \frac{3}{2}(\ell + \frac{1}{2}) - 
\frac{9}{4 \ell}) \varphi_3^2 \varphi_4 
+ \big ( 2(\ell + \frac{1}{2}) - \frac{3}{2 \ell}\big ) \varphi_3 \varphi_4 + \frac{9}{4} \varphi_3 \varphi_4^2 + 
\frac{1}{2 \ell} \varphi_3 \varphi_4^3 \Bigg) \,.
\end{multline}

\end{document}